\documentclass[11pt]{llncs}

\usepackage{latexsym}
\usepackage{graphicx}
\usepackage{ascmac}
\usepackage{color}

\renewcommand{\date}{\today}

\makeatletter
\spn@wtheorem{observation}{Observation}{\itshape}{\rmfamily}
\makeatother

\newcommand{\nop}[1]{}

\newcommand{\refA}[1]{\textcolor{black}{#1}}
\newcommand{\refB}[1]{\textcolor{black}{#1}}

\makeatletter
\leftmargin=\leftmargini
\labelwidth=\leftmargini
\advance \labelwidth by -\labelsep
\def\@listI{%
  \leftmargin=\leftmargini
  \parsep=\z@
  \topsep=2pt plus 2pt minus 2pt
  \itemsep=0pt plus 2pt\relax}%
\let \@listi=\@listI
\@listi

\def\@listii{%
  \leftmargin=\leftmarginii
  \labelwidth=\leftmarginii
  \advance \labelwidth by -\labelsep
  \topsep=2pt plus 1pt minus 1pt
  \parsep=\z@
  \itemsep=0pt plus 1pt\relax}%
\makeatother

\title{The complexity of UNO}

\author{%
   Erik D.~Demaine\inst{1} \and %
   Martin L.~Demaine\inst{1} \and %
   Nicholas J.~A.~Harvey\inst{2} \and 
   Ryuhei Uehara\inst{3} \and \\%
   Takeaki {\rm Uno}\inst{4} \and
   Yushi {\rm Uno}\inst{5}
}
\institute{
    MIT Computer Science and Artificial Intelligence Laboratory, 
    32 Vassar St., Cambridge, MA 02139, USA.
    \email{edemaine@mit.edu}, 
    \email{mdemaine@mit.edu}
\and
Department of Computer Science, Faculty of Science, 
The University of British Columbia, 
2329 West Mall, Vancouver, B.C., V6T 1Z4 Canada. 
\email{nichhar@cs.ubc.ca}
\and
    School of Information Science, JAIST, 
    Asahidai 1-1, Nomi, Ishikawa 923-1292, Japan. 
    \email{uehara@jaist.ac.jp}
\and
    National Institute of Informatics, 
    2-1-2 Hitotsubashi, Chiyoda-ku, Tokyo 101-8430, Japan. 
    \email{uno@nii.jp}
\and
    Graduate School of Science, Osaka Prefecture University, 
    1-1 Gakuen-cho, Naka-ku, Sakai 599-8531, Japan. 
    \email{uno@mi.s.osakafu-u.ac.jp}
}

\begin{document}
\maketitle

%
%
%
%
%
%
%
%
%
%
%
%
%

\begin{abstract}
This paper investigates the popular card game
UNO\textsuperscript{\textregistered}
from the viewpoint of algorithmic combinatorial game theory.
We define simple and concise mathematical models for the game,
including both cooperative and uncooperative versions,
and analyze their computational complexity.
In particular, we prove that even a single-player version of UNO is NP-complete,
although some restricted cases are in P.
\refA{Surprisingly, we show that the uncooperative two-player version is also in~P.}
\end{abstract}



\renewcommand{\thefootnote}{\fnsymbol{footnote}}
\setcounter{footnote}{3}

\section{Introduction}

Puzzles and games are enjoyable not only as a pastime,
but even their theoretical analysis has long been a source
of enjoyment for both mathematicians and computer scientists 
\cite{G05,HD09}. 
In particular, much research activity has studied the computational complexity of puzzles and games;
that is, how hard or easy it is to obtain a solution to a puzzle
or to decide the winner or loser of a game \cite{D01,ET76,LS80}. 
Some games and puzzles of interest include Nim, Hex, Peg Solitaire,
Instant Insanity, Tetris, 
Geography, Amazons, Chess, Othello, Go, Poker, and so on. 
Recently, 
this field is called `algorithmic combinatorial game theory' 
\cite{D01} to distinguish it from games arising from other fields, 
especially classical economic game theory. 

In this paper, we study UNO\footnote{UNO\textsuperscript{\textregistered}
is a registered trademark of Mattel Inc.}, 
an American card game invented by Merle Robbins in 1971
that has become surprisingly popular world-wide.
To study this game from the viewpoint of algorithmic combinatorial game theory,
we first propose mathematical models of UNO and then
analyze their computational complexities. 
We show that even a single-player version of UNO is computationally 
intractable, while the problem becomes rather easy under certain restrictions. 
Table~\ref{summary} summarizes our results.

\begin{table}
  \centering
  \begin{tabular}{l|l|l}
    \bf Model and number of players & \bf Complexity & \bf Reference \\ \hline
    \textsc{Solitaire Uno} (\textsc{Uno-1}) & NP-complete & Theorem~\ref{uno1hardness} \\
    \textsc{Solitaire Uno} (\textsc{Uno-1}) with & polynomial & Theorem~\ref{Uno-1 P} \\
    {\quad} $O(1)$ colors or $O(1)$ numbers & & \\
    \textsc{Cooperative Uno-2} & NP-complete & Theorem~\ref{UNO2NPC} \\
    \textsc{Uncooperative Uno-2} & polynomial & Theorem~\ref{uncoopUNO2isP}
  \end{tabular}
\vspace*{0.5cm}
  \caption{Summary of our results.}
  \label{summary}
\end{table}

We organize this paper as follows.
Section 2 introduces two mathematical models of UNO and their variants, 
and also defines UNO graphs. 
Among those models, Section 3 focuses on a single-player version of UNO, 
and examines its complexity; and
Section 4 considers a two-player version of UNO, 
\refA{which turns out to be in P.} 
Finally, Section 5 concludes the paper.

\section{Preliminaries}

Combinatorial games are often categorized by several 
properties that arise in theoretical modelling. 
Typical classifications are, for example, whether it is 
multi-player or single-player, imperfect-information or perfect-information, 
cooperative or uncooperative, and so on \cite{D01,HD09}. 
A single-player game is automatically perfect-information and cooperative, 
and is usually called a puzzle.

\subsection{Game setting}
\label{subsec:game_settings}

The card game UNO
can be played by 2--10 players. 
At the beginning of the game, 
each player is dealt an equal number of cards.
Aside from some special cards called `action cards',
each (normal) card has a color and a number.
Players play in turn.
In each turn,
the player can \refB{play} one of his/her cards
by matching the card's color or number 
to the one \refB{played} immediately beforehand.
Alternatively, in particular when the player has no such card to play,
the player can draw a card from the draw pile, and then optionally
play that card if possible.
The objective of a single game is to be the first player 
to \refB{play} all the cards in one's hand before one's opponents. 
Thus, UNO is 
a (i) multi-player, (ii) imperfect-information, 
and (iii) uncooperative combinatorial game.
(See \ref{rule_of_uno} for more detailed rules of UNO.)

In reality, action cards make UNO more complicated 
and interesting, 
because they may prompt psychological strategies. 
However, for mathematical simplicity,
this paper concentrates on the most important aspect of the rules of UNO:
each card has two attributes, a color and a number, 
and that one can \refB{play} a card only if its color or number 
matches the card \refB{played} immediately before one's turn. 
Our mathematical models make the following assumptions:
(a) we do not take into account action cards nor the draw pile;
(b) all the cards dealt to and in the hand of any player are open 
during the game (i.e., perfect-information); 
(c) we do not necessarily assume 
that all the players have the same number of cards 
at the beginning of a game (unless otherwise stated); 
(d) every player acts rationally (e.g., 
no player is allowed to skip their turn intentionally); and 
(e) the first player can start the game 
by \refB{playing} any card he/she likes at hand,
with no constraint on matching color or number.

\subsection{Definitions and Notations}

An UNO card has two attributes called \emph{color} and \emph{number}, 
and in general, we define a \emph{card} to be a tuple $(x,y)$ $\in X\times Y$, 
where $X=\{1,2,\ldots ,c\}$ is a set of colors 
and $Y=\{1,2,\ldots ,b\}$ is a set of numbers. 
A finite number of \emph{players} $1,2,\ldots ,p$ $(\ge 1)$ can join an UNO game. 
At the beginning of a single game of UNO, 
each card of a set of $n$ cards $C$ is dealt to one player among $p$ players, 
i.e., each player $i$ is initially given a set $C_i$ of cards: 
$C_i=\{t_{i,1},\ldots ,t_{i,n_i}\}$ $(i=1,2,\ldots ,p)$. 
By definition, $\sum_{i=1}^{p}n_i=n$. 
Here, we assume that $C$ is a multiset; that is, 
there may be more than one card with the same color and the same number. 
We denote a card $(x,y)$ dealt to player $i$ by $(x,y)_i$. 
When the number of players is one, we omit the subscript 
without any confusion. 
Throughout the paper, we assume without loss of generality 
that player 1 is the first to play, 
and players $1,2,\ldots ,p$ play in turn in this order. 

Player $i$ can \refB{\emph{play}} (or \emph{discard}) exactly one card 
currently at hand in his/her turn 
if the color or the number of the card is equal to that of the card 
\refB{played} immediately before player $i$. 
In other words, 
we say that a card $t'=(x',y')_{i'}$ can be \refB{played} 
immediately after a card $t=(x,y)_i$ 
if and only if $((x'=x) \vee (y'=y))\wedge (i'=i+1\ \mbox{(mod $p$)})$. 
We also say that a card $t'$ \emph{matches} a card $t$ 
when $t'$ can be \refB{played} after $t$. 
A \refB{played} card is removed from the set of cards in the player's hand.
A \refB{\emph{playing}} (or \emph{discarding}) \emph{sequence} (\emph{of cards}) 
of a card set $C$ is a sequence of cards $(t_{s_1},\ldots ,t_{s_k})$ 
such that $t_{s_i}\in C$ and $t_{s_j}\in C\setminus \{t_{s_i}\mid i<j\}$ 
for each $j\in \{1,\ldots ,k\}$. 
A \refB{playing} sequence $(t_{s_1},\ldots ,t_{s_k})$ is \emph{feasible} 
if $t_{s_{j+1}}$ matches $t_{s_j}$ for $j=1,2,\ldots ,k-1$. 

In our mathematical models of UNO, 
we specify the problems by four parameters: 
the number of players $p$, the number of total cards $n$, 
the number of colors $c$, and the number of numbers $b$. 
The two values $c$ and $b$ are assumed to be unbounded 
unless otherwise stated.

\subsection{Models}

We now define two different versions of UNO;
one is cooperative and the other is uncooperative. 

\begin{quote}
{\sc Uncooperative Uno}
\begin{description}
\item[Instance:] the number of players $p$, 
and player $i$'s card set $C_i$ 
with $c$ colors and $b$ numbers.
\item[Question:]
determine the first loser; 
i.e., the player that cannot play one's card any more 
in spite that his/her hand is not empty. 
\end{description}
\end{quote}
We refer to {\sc Uncooperative Uno} with $p$ players 
as {\sc Uncooperative Uno-$p$}. 
This problem setting makes sense only if $p\ge 2$ 
because UNO played by a single player 
automatically becomes cooperative. 

\begin{quote}
{\sc Cooperative Uno}
\begin{description}
\item[Instance:] the number of players $p$, 
player $i$'s card set $C_i$ 
with $c$ colors and $b$ numbers.
\item[Question:]
can all the players make player 1 win, 
i.e., make player 1's card set empty before any of the other players 
become finished? 
\end{description}
\end{quote}
We abbreviate {\sc Cooperative Uno} played by $p$ players 
as {\sc Cooperative Uno-$p$}, or simply as {\sc Uno-$p$}. 
This problem setting makes sense if the number of players $p$ 
is greater than or equal to 1. 
In {\sc Uncooperative/Cooperative Uno}, 
when the number of players is given by a constant, such as {\sc Uno-2}, 
this implies that $p$ is no longer a part of the input of the problem. 
In addition to the assumptions (a)--(e) on game settings 
described in Section~\ref{subsec:game_settings}, 
we set one additional assumption which changes depending 
on whether the game is cooperative or uncooperative: 
any player who cannot \refB{play} any card at hand 
(f1) skips their turn 
but still remains in the game and waits for the next turn 
in cooperative games, and 
(f2) is a loser in uncooperative games.

We define an \emph{UNO-$p$ graph} as the directed graph 
representing the `match' relationship among cards 
in the entire card set $C$. 
More precisely, a vertex corresponds to a card, and there is a directed arc 
from vertex $u$ to $v$ if and only if the corresponding card $t_v$ 
matches (can be \refB{played} immediately after) $t_u$. 
Let us consider an UNO-1 graph, i.e., an UNO-$p$ graph in the case 
that the number of players is $p=1$. 
In this case, a card $t'$ matches $t$ if and only if $t$ matches $t'$;
that is, the `match' relation is symmetric. 
Thus UNO-1 graphs can be viewed as undirected. 
For an UNO-2 graph, 
a card $t'=(x',y')_2$ matches $t=(x,y)_1$ if and only if $t$ matches $t'$, 
and therefore UNO-2 graphs also become undirected. 
Furthermore, because a player cannot play consecutively 
when the number of players is $p\ge 2$, an UNO-2 graph becomes bipartite. 
In general, because $n$ cards of a card set $C$ are dealt to $p$ players 
at the beginning of a single UNO game
(i.e., $C$ is partitioned into $C_i=\{(x,y)_i\}$),
UNO-$p$ graphs are (restricted) $p$-partite graphs
whose partite sets correspond to $C_i$.

\section{Cooperative UNO}
\label{sec:cooperative}

In this section, we focus on the cooperative version of UNO, 
and discuss its complexity when the number of players is two or one. 

\subsection{Two-player case}

We first show that {\sc Uno-2} is intractable. 

\begin{theorem}
\label{UNO2NPC}
{\sc Uno-2} is NP-complete. 
\end{theorem}

\begin{figure}[hbt]
\centering
\scalebox{0.75}{\input{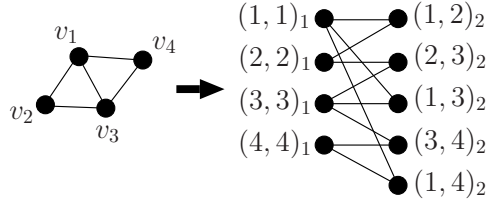}}
\caption{Reduction from {\sc HP} to {\sc Uno-2}.}
\label{hp2unotwo}
\end{figure}

\begin{proof}
By reduction from {\sc Hamiltonian Path} (HP). 
An instance of HP 
is given by an undirected graph $G$. 
The problem asks whether there is a Hamiltonian path in $G$, 
and it is known to be NP-complete \cite{GJ79}. 
Here, we assume without loss of generality 
that $G$ is connected and is not a tree, 
and hence that $|V(G)|\le |E(G)|$. 

We transform an instance of HP 
into an instance of {\sc Uno-2} as follows. 
Let $C_1$ and $C_2$ be the card set of players 1 and 2, respectively. 
We use $C_1$ to represent vertices of $G$, and $C_2$ to represent edges of $G$,
by defining $C_1=\{(i,i)\mid v_i\in V(G)\}$
and $C_2=\{(i,j)\mid \{v_i,v_j\}\in E(G)\}$.
The UNO-2 graph $G'$ is thus the vertex--edge incidence graph of~$G$;
see Figure~\ref{hp2unotwo}.
Now we show that the answer to an instance of HP is yes 
if and only if the answer to the corresponding instance of {\sc Uno-2} is yes. 
If there is a Hamiltonian path, say $P=(v_{i_1},v_{i_2},\ldots ,v_{i_n})$, 
in the instance graph of HP, 
then there is a feasible \refB{playing} sequence 
$((i_1,i_1)_1,(i_1,i_2)_2,(i_2,i_2)_1,\ldots ,$ 
$(i_{n-1},i_{n-1})_1,$ $(i_{n-1},i_n)_2,$ $(i_n,i_n)_1)$
in which player 1 discards all cards before player 2 does. 
Conversely, if there is a feasible \refB{playing} sequence 
$((i_1,i_1)_1,(i_1,i_2)_2,(i_2,i_2)_1,\ldots ,$ 
$(i_{n-1},i_{n-1})_1,$ $(i_{n-1},i_n)_2,$ $(i_n,i_n)_1)$,
it visits all vertices in $C_1$ of $G'$ exactly once, 
and thus the corresponding sequence of vertices 
$(v_{i_1},$ $v_{i_2},$ $\ldots ,$ $v_{i_n})$ 
is a simple path visiting all the vertices in $V(G)$ exactly once,
i.e., a Hamiltonian path in $G$. 

The size of the {\sc Uno-2} instance is proportional to $|C_1|+|C_2|$. 
Because $|C_1|=|V(G)|$ and $|C_2|=|E(G)|$, 
the reduction has polynomial size in $|V(G)|+|E(G)|$, 
which is the input size of the {\sc HP} instance.

Finally we argue that {\sc Uno-2} is in NP.
For any {\sc Uno-2} instance, a playing sequence has length at most equal
to the number $n$ of cards.  Thus we can nondeterministically guess a
playing sequence, and verify it in polynomial time.
\qed
\end{proof}

\begin{corollary}
\label{UNO2NPCeq}
{\sc Uno-2} is NP-complete 
even when the number of cards of the two players are equal. 
\end{corollary}

\begin{proof}
By reduction from {\sc Hamiltonian Path} with specified starting vertex~$v_i$, 
which is known to be NP-complete \cite{GJ79}. 
We start from the same reduction in the proof of Theorem~\ref{UNO2NPC}.
As in that proof, we can assume $|C_1|\le |C_2|$ without loss of generality. 
If $|C_1|=|C_2|$, we are done. 
If $|C_1|<|C_2|$, we add $|C_2|-|C_1|$ cards $(n+2,n+2)$ 
and a single card $(n+2,n+1)$ to $C_1$, 
and add a single card $(i,n+1)$ to $C_2$.
After this change, $|C_1|=|C_2|$.
If there is a Hamiltonian path starting at $v_i$,
player~1 can now win by repeatedly playing $(n+2,n+2)$,
allowing player~2 to skip their turn as they have no matching card,
until player~1 exhausts all $(n+2,n+2)$ cards
and then plays $(n+2,n+1)$, to which player~2 has a unique response
$(i,n+1)$, and the play proceeds as in the previous reduction
(but starting at vertex~$v_i$).
In fact, the only valid play sequences will either start in this way,
or end symmetrically with $(i,n+1)_1, (n+2,n+1)_2, (n+2,n+2)_1, \dots,
(n+2,n+2)_1$, so play sequences are equivalent to Hamiltonian paths starting
(or ending) at~$v_i$.
\qed
\end{proof}

\subsection{Single-player, intractable case}

In the single-player case, 
the cooperative and uncooperative versions of UNO
become equivalent. 
We redefine this setting as follows. 

\begin{quote}
{\sc Uno-1 (Solitaire Uno)}
\begin{description}
\item[Instance:] a set $C$ of $n$ cards $(x_i,y_i)$ $(i=1,2,\ldots ,n)$, 
where $x_i\in \{1,2,\ldots ,c\}$ and $y_i\in \{1,2,\ldots ,b\}$.
\item[Question:] determine whether the player can \refB{play} all the cards. 
\end{description}
\end{quote}

\begin{figure}[hbt]
\begin{minipage}{1.0\linewidth}
\centering
\scalebox{0.75}{\input{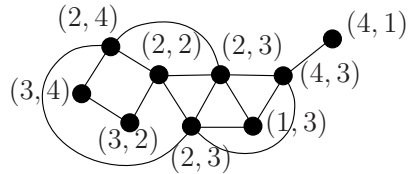}}
\caption{An example of UNO-$1$ graph.}
\label{uno1graph}
\end{minipage}
\end{figure}

\begin{example}
\label{single}
Let the card set $C$ for player 1 be given by 
$C=\{(1,3)$, $(2,2)$, $(2,3)$, $(2,3)$, $(2,4)$, $(3,2)$, $(3,4)$, $(4,1)$, 
$(4,3)\}$. 
Then, a feasible \refB{playing} sequence using all the cards is 
$((1,3)$, $(2,3)$, $(2,4)$, $(3,4)$, $(3,2)$, $(2,2)$, $(2,3)$, $(4,3)$, 
$(4,1))$ in this order, and the answer is yes. 
Figure~\ref{uno1graph} shows the corresponding UNO-$1$ graph.
\end{example}

We investigate here some basic properties of UNO-1 graphs. 
In UNO-1 graphs, all the vertices whose corresponding cards 
have either the same color or the same number form a clique. 
The \emph{line graph} $L(G)$ of a given graph $G$ is the graph 
whose vertices are edges of $G$ and $\{e,e'\}\in E(L(G))$ 
for $e,e'\in V(L(G)) = E(G)$
if and only if $e$ and $e'$ share an endpoint in~$G$. 
\refA{Because an UNO card is an ordered pair of a color and a number, 
UNO cards correspond to the edge set of a bipartite graph 
whose partite sets are colors and numbers. 
Then an UNO-1 graph represents the adjacency of edges 
(corresponding to cards) of a bipartite graph. 
These arguments lead the following fact.} 

\begin{observation}
\label{claw-free}
A graph is UNO-1 if and only if it is a line graph of a bipartite graph. 
\end{observation}

{\sc Uno-1} is essentially equivalent to finding a Hamiltonian path in an
UNO-1 graph, which we now know are line graphs of bipartite graphs.
The following fact is known. 

\begin{theorem}{\rm \cite{LW93}}
{\sc Hamiltonian Path} for line graphs of bipartite graphs is NP-complete. 
\end{theorem}

Therefore, as a corollary of this theorem, 
we see that UNO is hard even for a single player. 

\begin{theorem}
\label{uno1hardness}
{\sc Uno-1} is NP-complete. 
\end{theorem}

For the sake of being self-contained and complete,
we give a direct and concise proof of Theorem~\ref{uno1hardness}.
By contrast, the proof in \cite{LW93} further depends on \cite{B81}. 

\medskip
\begin{proof}
A cubic graph is a graph in which every vertex has degree~$3$. 
We reduce from {\sc Hamiltonian Path} for cubic graphs ({\sc HP-C}), 
which is known to be NP-complete \cite{GJT76}.

Consider an instance $G$ of {\sc HP-C}.
We transform $G$ into a graph $G'$, where 
\[
\begin{array}{l}
V(G')~=~\{(x,e)\mid x\in V(G),e=\{x,y\}\in E(G)\}, \\
E(G')~=~\{((x,e),(y,e))\mid e=\{x,y\}\in E(G)\}
~\cup~ \{((x,e_i),(x,e_j))\mid e_i\neq e_j\}. 
\end{array}
\]
This transformation splits any vertex $x\in V(G)$
into three new vertices $(x,e_1),$ $(x,e_2),$ $(x,e_3)$ to form a clique (triangle), 
while each edge $e_i$ $(i=1,2,3)$ incident to $x$ 
becomes incident to a new vertex $(x,e_i)$. 
Figure~\ref{triangle_gadget} illustrates this \emph{node gadget}.
Then we prepare the card set $C$ of the player of {\sc Uno-1} 
to be the set $V(G')$, 
where the color and the number of $(x,e)$ are $x$ and $e$, respectively. 
We can easily confirm that there is an edge $e=(t,t')$ in $G'$ 
if and only if $t$ and $t'$ match.
Thus, $G'$ is the corresponding UNO-1 graph for card set $C$. 
Now it suffices to show that there is a Hamiltonian path in $G$
if and only if there is a Hamiltonian path in $G'$. 

\begin{figure}[hbt]
\centering
\scalebox{0.75}{\input{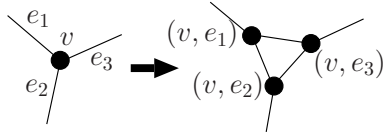}}
\caption{A node gadget splits a vertex into three vertices to form a triangle.}
\label{triangle_gadget}
\end{figure}

First suppose that there is a Hamiltonian path in $G$, say $P=(v_{i_1},\ldots ,v_{i_n})$. 
We construct a Hamiltonian path $P'$ in $G'$ from $P$ as follows. 
Let $v_{i_{j-1}},v_{i_j},v_{i_{j+1}}$ be three consecutive vertices in $P$,
in this order.
Let $e_1=\{v_{i_{j-1}},v_{i_j}\}$, $e_2=\{v_{i_j},v_{i_{j+1}}\}$,
and let $e_3=\{v_{i_j},v_{i_k}\}$ be the unique edge with $k \not\in \{j-1,j+1\}$. 
Then we replace these three vertices 
by the sequence of vertices $(v_{i_{j-1}},e_1)$, $(v_{i_j},e_1)$, 
$(v_{i_j},e_3)$, $(v_{i_j},e_2)$, $(v_{i_{j+1}},e_2)$ in $G'$ 
to form a subpath in $P'$. 
For the starting two vertices $v_{i_1}$ and $v_{i_2}$, 
we replace them by the sequence of vertices $(v_{i_1},e_1)$, $(v_{i_1},e_2)$,
$(v_{i_1},\{v_{i_1},v_{i_2}\})$, $(v_{i_2},\{v_{i_1},v_{i_2}\})$,
where $e_1$ and $e_2$ are the two edges incident to $v_{i_1}$ other than $\{v_{i_1},v_{i_2}\}$.
The final two vertices are handled similarly.
One may confirm that the resulting sequence of vertices $P'$ in $G'$ 
form a Hamiltonian path. 

For the converse, suppose that there is a Hamiltonian path $P'$ in UNO-1 graph~$G'$.
If $P'$ visits $(v,e_i)$ $(i=1,2,3)$ consecutively in any order 
(call this property \emph{consecutiveness}) for any~$v$,
as shown in Figure~\ref{possible_tours} (a1) or (a2),
then $P'$ can be transformed into a Hamiltonian path $P$ in $G$ 
in an obvious way. 
Suppose, however, that a Hamiltonian path $P'$ in $G'$ does not 
visit $(v,e_i)$ $(i=1,2,3)$ consecutively. 
It suffices to show that such $P'$ can be transformed into another path 
satisfying consecutiveness. 
There are two possible cases,
as shown in Figure~\ref{possible_tours} (b') and (c').
In both of these cases, at least one endpoint of $P'$ has the form $(v,e_i)$,
and we can resolve this nonconsecutiveness by visiting $(v,e_i)$ when we
visit the other $(v,e_j)$'s, instead of as an endpoint of the path,
as in Figure~\ref{possible_tours} (b) and (c).

\begin{figure}[hbt]
\centering
\scalebox{0.75}{\input{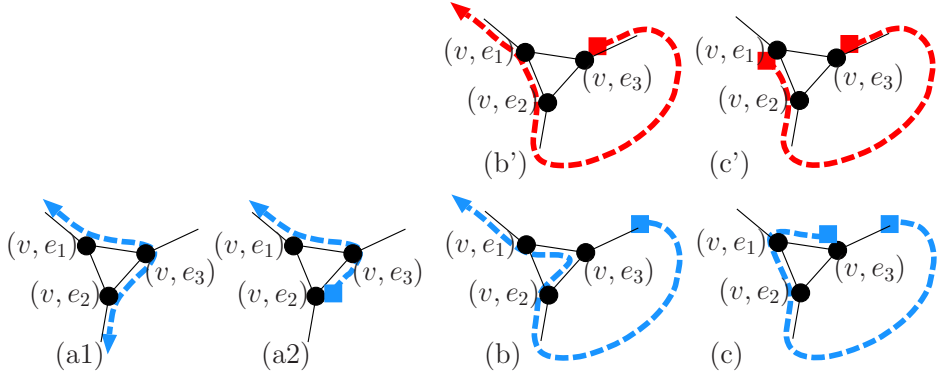}}
\caption{Possible tours passing through a node gadget. 
Each dotted line is a part of a tour, 
and a square denotes either end of a tour. } 
\label{possible_tours}
\end{figure}

The size of the reduced instance is proportional to the size of the
{\sc HP-C} instance.  {\sc Uno-1} is in NP as in the proof of Theorem~\ref{UNO2NPC}.
Thus, the proof is complete. 
\qed
\end{proof}

\subsection{Single-player, tractable case}

In the remainder of this section, 
we show that the generally intractable problem {\sc Uno-1} becomes tractable 
if the number of colors $c$ is bounded by a constant. 
Our algorithm is based on dynamic programming (DP).

\begin{figure}[hbt]
\centering
\scalebox{0.74}{\input{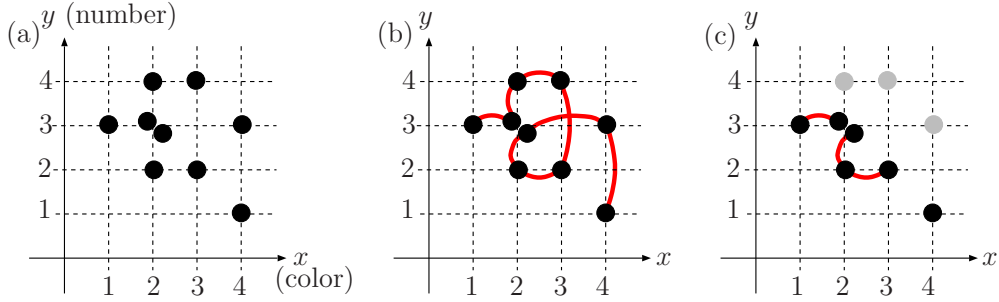}}
\caption{(a) Geometric view of an UNO-1 graph, 
where all the edges are omitted, 
(b) a Hamiltonian path in the UNO-1 graph, and 
(c) a set of subpaths in the subgraph of the UNO-1 graph induced 
by the first six points; it shows $h_{\{1,2\}}=1$, $v_{(2,3)}=1$ 
and $d_{\{4,4\}}=1$.}
\label{geom_view}
\end{figure}

First we introduce a geometric view of UNO-1 graphs. 
Because an UNO card $(x,y)$ is an ordered pair of integer values 
representing its color and number, 
the card can be viewed as an (integer) lattice point in the 2-dimensional 
square lattice plane.
Then an UNO-1 graph is a set of points in that plane, 
where all the points with the same $x$- or $y$-coordinate form a clique. 
We call this interpretation the \emph{geometric view} of UNO-1 graphs. 
Figure~\ref{geom_view}(a) illustrates
the geometric view of the instance from Example~\ref{single}.
Now the {\sc Uno-1} problem
asks, for a given set of points in the plane 
and starting and ending at appropriate different points, 
whether all the points can be visited exactly once 
by a path that makes only axis-parallel moves;
see Figure~\ref{geom_view}(b). 

\paragraph{Approach}
Let $C$ be a set of $n$ points and $G$ be an UNO-1 graph defined by $C$. 
Then a subgraph $P$ forms a Hamiltonian path if and only if 
it is a single path that spans $G$. 
Suppose a subgraph $P$ is a spanning path of $G$. 
If we consider a subset $C'$ 
of the point set $C$, 
then $P[C']$ (the subgraph of $P$ induced by $C'$)
is a set of subpaths that spans $G[C']$ (Figure~\ref{geom_view}(c)). 
We count and maintain the number of sets of subpaths 
by classifying subpaths into three classes
according to the types of their two endpoints. 

Starting with the empty set of points, 
the DP proceeds by adding a new point according to a fixed order 
by updating the number of sets of subpaths iteratively. 
Finally, when the set of points grows to $C$, 
we can confirm the existence of a Hamiltonian path in $G$ 
by checking the number of sets of subpaths 
consisting of a single subpath (without isolated vertices). 
We remark that, throughout this DP, 
an isolated vertex is regarded for convenience 
as being a (virtual) path starting and ending at itself.

\paragraph{Mechanism}
To specify the order in which points get added into the DP, 
define the total order $\prec$ as lexical order of inverted $(y,x)$ tuples.
More precisely,
if $x(t)$ and $y(t)$ denote the $x$- and $y$-coordinates of a point~$t$, 
and $t$ and $t'$ are two points in $C$, then
define $t\prec t'$ if $y(t)<y(t')$ or $[y(t)=y(t')\wedge x(t)<x(t')]$. 
Because $C$ is a multiset, we must also define $t \prec t'$
when $t=t'$; in this case, break ties arbitrary. 
Let $t_1,t_2,\ldots,t_n$ denote the $n$ points in $C$
in increasing order by $\prec$.
Define prefix subsets $C_{\ell}=\{t_i\mid 1\le i\le \ell\}$. 

At step~$\ell$ of the DP, we consider adding
a new point $t_{\ell}$ $=(x(t_{\ell}),y(t_{\ell}))$ 
to the existing prefix $C_{\ell -1}$. 
Point $t_\ell$ must be added to two, one, or zero endpoints of different
subpaths to form a new set of subpaths. 

Let $\mathcal{P}(\ell)$ be a family of sets of subpaths 
spanning $G[C_{\ell}]$. 
(Recall that we regard an isolated vertex as being a path spanning itself.) 
Then we classify subpaths in a set of subpaths $\mathcal{P}$ 
$\in \mathcal{P}(\ell)$ in the following manner.
For any subpath $P$ $\in \mathcal{P}$ and the $y$-coordinates 
of its two endpoints, one of the following cases holds:
(i) both equal $y(t_{\ell})$ (type-h), 
(ii) exactly one equals $y(t_{\ell})$ (type-v), or 
(iii) none equals $y(t_{\ell})$ (type-d) holds. 
We count the number of these three types of subpaths in $\mathcal{P}$ 
further by classifying them by the $x$-coordinates of their endpoints. 
(Notice that type-h and type-d are symmetric 
in the sense that paths of these types do not distinguish their end points, 
but type-v is not.) 
For this purpose, we prepare some subscript sets: 
a set of subscripts $K=\{1,2,\ldots ,c\}$, 
sets of unordered pairs of subscripts 
$I={K\choose 2}$ and $I^+=I\cup \{\{i,i\}\mid i\in K\}$, and 
sets of ordered pairs of subscripts 
$J=K\times K$ and $J^-=J-\{(i,i)\mid i\in K\}$. 

We introduce the following parameters $h$, $v$, and $d$ 
to count the number of subpaths in $\mathcal{P}$ $(\in \mathcal{P}({\ell}))$ 
(see Figure~\ref{geom_view}(c)): 
\[
\arraycolsep=0.5pt
\begin{array}{lcl}
h_{\{i,i'\}} &=& \mbox{\#subpaths in $\mathcal{P}$ 
with endpoints $(x_i,y(t_{\ell}))$ and $(x_{i'},y(t_{\ell}))$ 
for $\{i,i'\}$ $\!\in\! I^+$}, \\
v_{(i,i')} &=& \mbox{\#subpaths in $\mathcal{P}$ 
with endpoints $(x_i,y(t_{\ell}))$ and $(x_{i'},y')$}
\\&&
\multicolumn{1}{r}{\mbox{for $(i,i')$ $\!\in\! J$ and $y'\!\!<\!y(t_{\ell})$},} \\
d_{\{i,i'\}} &=& \mbox{\#subpaths in $\mathcal{P}$ with endpoints 
$(x_i,y')$ and $(x_{i'},y'')$}
\\&&
\multicolumn{1}{r}{\mbox{for $\{i,i'\}$ $\!\in\! I^+$ and $y',y''\!\!<\!y(t_{\ell})$}.}
\end{array}
\]
Then we define a $(2|I^+|+|J|)$-dimensional vector 
$z(\mathcal{P})$ for a set of subpaths $\mathcal{P}$ 
$(\in\mathcal{P(\ell)})$ as 
$z(\mathcal{P})$ $=(\vec{h};\vec{v};\vec{d})$ 
$=(\langle h_{\{1,1\}},\ldots ,$ $h_{\{1,c\}},h_{\{2,2\}},\ldots ,$ 
$h_{\{2,c\}},h_{\{3,3\}},\ldots ,$ $h_{\{c,c\}}\rangle$; 
$\langle v_{(1,1)},\ldots ,$ $v_{(1,c)}$, $v_{(2,1)}$, $v_{(2,2)}$, 
$\ldots$, $v_{(2,c)},v_{(3,1)},\ldots ,$ $v_{(c,c)}\rangle$; 
$\langle d_{\{1,1\}},\ldots ,$ $d_{\{1,c\}},$ $d_{\{2,2\}},\ldots ,$ 
$d_{\{2,c\}},d_{\{3,3\}},\ldots ,$ $d_{\{c,c\}}\rangle)$. 
Finally, for a given vector $(\vec{h};\vec{v};\vec{d})$, 
we define the number of sets $\mathcal{P}$ satisfying 
$z(\mathcal{P})$ $=(\vec{h};\vec{v};\vec{d})$ 
in a family $\mathcal{P}(\ell)$ by $f(\ell,(\vec{h};\vec{v};\vec{d}))$, 
i.e., $f(\ell,(\vec{h};\vec{v};\vec{d}))$ 
$=\bigl|\{\,\mathcal{P}$ $\mid \mathcal{P}\in \mathcal{P}(\ell)$, 
$z(\mathcal{P})=(\vec{h};\vec{v};\vec{d})\,\}\bigr|$. 
Now the objective of the DP is to determine whether
there exists a vector $(\vec{h};\vec{v};\vec{d})$ 
such that $f(n,$ $(\vec{h};\vec{v};\vec{d}))$ $\ge 1$, 
where all the elements in $\vec{h}$, $\vec{v}$, and $\vec{d}$ are 0, 
except for exactly one element that is 1.

\paragraph{Recursion} 
As we explained, 
the DP proceeds by adding a new point $t_{\ell}$ to $C_{\ell-1}$. 
When $t_{\ell}$ is added, it is connected to 0, 1 or 2 endpoints 
of existing different paths, 
where each endpoint has $y(t_{\ell})$ or $x(t_{\ell})$ in its coordinate. 
The recursion of the DP is described just by summing up 
all possible combinations of these patterns. 
We divide into three cases, one of which has two subcases: 
(a) a set of base cases; 
(b) a case in which $t_{\ell}$ is added as the first point 
whose $y$-coordinate is $y(t_{\ell})$, and 
(b1) as an isolated vertex, or 
(b2) as to be connected to an existing path; 
(c) all other cases.

\begin{figure}[hbt]
\centering
  \begin{center}
\scalebox{0.75}{\input{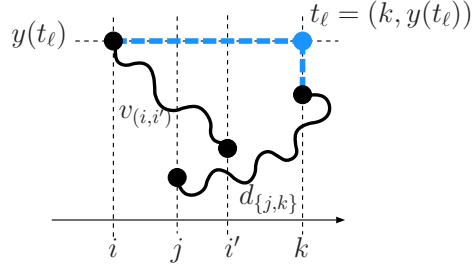}}
  \end{center}
\caption{An example case of the DP.}
\label{dp_example}
\end{figure}

Now we can give the DP formula 
for computing $f(\ell;(\vec{h};\vec{v};\vec{d}))$. 
For its correctness, however, to avoid complication 
we just explain the idea of the DP in Figure~\ref{dp_example} 
by illustrating one of the cases appearing in the DP. 
In this example, consider a subpath in a graph induced by $C_{\ell}$ 
whose two endpoints have $x_{i'}$ and $x_{j}$ in their $x$-coordinates. 
It will be counted in $d_{\{j,i'\}}$. 
Then this subpath can be generated 
by adding the point $t_{\ell}$ to connect to two paths 
in a graph induced by $C_{\ell-1}$:
the one whose one endpoint is $(x_i,y(t_{\ell}))$ (counted in $v_{(i,i')}$), 
and the other one whose one endpoint is $(k,y)$ $(y<y(t_{\ell}))$ 
(counted in $d_{\{j,k\}}$). 
The number of such paths is the sum of those 
for all the combinations of $i$, $i'$, and~$j$. 

To describe the recursion, we need some notation.
Let $\vec{h}<\vec{0}$ mean that $h_{\{i,i'\}}<0$ holds
for some subscript $\{i,i'\}$ 
(and similarly for $\vec{v}$ and $\vec{d}$).
Let $k=x(t_{\ell})$.
Let $\delta_{ij}$ be Kronecker's delta function:
\[ \delta_{ij} = \left\{\begin{array}{ll}
      0 & \mbox{if $i \neq j$}, \\
      1 & \mbox{if $i = j$}.
   \end{array}\right. \]

The overall recursion divides into the cases as follows:
\[
f(\ell;(\vec{h};\vec{v};\vec{d}))=
\left\{
\begin{array}{llp{2in}}
0 & \mbox{(a)} & if $\ell\!=\!0$ or $\vec{h}\!<\!\vec{0}$ 
or $\vec{v}\!<\!\vec{0}$ or $\vec{d}\!<\!\vec{0}$, \\
f_{b1}(\ell;(\vec{h};\vec{v};\vec{d}))
& \mbox{(b1)} & if $\ell\!>\!0$, $y(t_{\ell})\!>\!y(t_{\ell-1})$, 
$\vec{h}\!=\!\langle 0\ldots h_{\{k,k\}}\!=\!1\ldots 0\rangle, 
\vec{v}\!=\!\vec{0}$,\\
f_{b1}(\ell;(\vec{h};\vec{v};\vec{d}))
& \mbox{(b2)} & if $\ell\!>\!0$, $y(t_{\ell})\!>\!y(t_{\ell-1})$, 
$\vec{h}\!=\!\vec{0}, \vec{v}\!=\!\langle 0\ldots v_{(k,j)}\!=\!1\ldots 0\rangle$, \\
f_c(\ell;(\vec{h};\vec{v};\vec{d}))
 & \mbox{(c)} & otherwise.
\end{array}
\right.
\]

The recursion for case (b1) is 
\[
\arraycolsep=3pt
\begin{array}{llcl}
& f_{b1}(\ell;(\vec{h};\vec{v};\vec{d})) \\
= & \sum_{\vec{h'},\vec{v'},\vec{d'}} \bigl\{ f(\ell\!-\!1; 
(\vec{h'};\vec{v'};\vec{d'})\mid \{i,i'\}\!\in\! I^+,
\\&\multicolumn{1}{r}{
d_{\{i,i'\}}\!=\!h'_{\{i,i'\}}\!+\!d'_{\{i,i'\}}\!+\!v'_{(i,i')}\!+\!(1\!-\!\delta_{ii'})v'_{(i'i)}} & (\{i,i'\}\!\neq\!\{k,j\}) 
\\&\multicolumn{1}{r}{
d_{\{i,i'\}}\!=\!(h'_{\{i,i'\}}\!-\!1)\!+\!d'_{\{i,i'\}}\!+\!v'_{(i,i')}\!+\!(1\!-\!\delta_{ii'})v'_{(i'i)}} & (\{i,i'\}\!=\!\{k,j\}) & \}
\end{array}
\]

The recursion for case (b2) is 
\[
\arraycolsep=3pt
\begin{array}{llcl}
& f_{b2}(\ell;(\vec{h};\vec{v};\vec{d})) \\
= & \sum_{\vec{h'},\vec{v'},\vec{d'}} \bigl\{ f(\ell\!-\!1; 
(\vec{h'};\vec{v'};\vec{d'})\mid  \{i,i'\}\!\in\! I^+,
\\&\multicolumn{1}{r}{
d_{\{i,i'\}}\!=\!h'_{\{i,i'\}}\!+\!d'_{\{i,i'\}}\!+\!v'_{(i,i')}\!+\!(1\!-\!\delta_{ii'})v'_{(i'i)}} & (\{i,i'\}\!\neq\!\{k,j\})
\\&\multicolumn{1}{r}{
d_{\{i,i'\}}\!=\!(h'_{\{i,i'\}}\!-\!1)\!+\!d'_{\{i,i'\}}\!+\!v'_{(i,i')}\!+\!(1\!-\!\delta_{ii'})v'_{(i'i)}} & (\{i,i'\}\!=\!\{k,j\}) & \} \\ 
+ & \sum_{\vec{h'},\vec{v'},\vec{d'}} \bigl\{ f(\ell\!-\!1; 
(\vec{h'};\vec{v'};\vec{d'})\mid \{i,i'\}\!\in\! I^+,
\\&\multicolumn{1}{r}{
d_{\{i,i'\}}\!=\!h'_{\{i,i'\}}\!+\!d'_{\{i,i'\}}\!+\!v'_{(i,i')}\!+\!(1\!-\!\delta_{ii'})v'_{(i'i)}} & (\{i,i'\}\!\neq\!\{k,j\})
\\&\multicolumn{1}{r}{
d_{\{i,i'\}}\!=\!h'_{\{i,i'\}}\!+\!d'_{\{i,i'\}}\!+\!(v'_{(i,i')}\!+\!(1\!-\!\delta_{ii'})v'_{(i'i)}\!-\!1)} & (\{i,i'\}\!=\!\{k,j\}) & \} \\ 
+ & \sum_{\vec{h'},\vec{v'},\vec{d'}} \bigl\{ f(\ell\!-\!1; 
(\vec{h'};\vec{v'};\vec{d'})\mid \{i,i'\}\!\in\! I^+,
\\&\multicolumn{1}{r}{
d_{\{i,i'\}}\!=\!h'_{\{i,i'\}}\!+\!d'_{\{i,i'\}}\!+\!v'_{(i,i')}\!+\!(1\!-\!\delta_{ii'})v'_{(i'i)}} & (\{i,i'\}\!\neq\!\{k,j\}) 
\\&\multicolumn{1}{r}{
d_{\{i,i'\}}\!=\!h'_{\{i,i'\}}\!+\!(d'_{\{i,i'\}}\!-\!1)\!+\!v'_{(i,i')}\!+\!(1\!-\!\delta_{ii'})v'_{(i'i)}} & (\{i,i'\}\!=\!\{k,j\}) & \}, 
\end{array}
\]

The recursion for case (c) is 
\[
\arraycolsep=3pt
\begin{array}{ll}
& f_c(\ell;(\vec{h};\vec{v};\vec{d})) \\
= & f(\ell;(\langle\ldots h_{\{k,k\}}\!-\!1\ldots\rangle;\vec{v};\vec{d})) \\
+ & \sum_{\{i,i'\}\in I}f(\ell;(\langle\ldots h_{\{i,i'\}}\!+\!1\ldots h_{\{i,k\}}\!-\!1\ldots\rangle;\vec{v};\vec{d})) \\
+ & \sum_{\{i,i'\}\in I}f(\ell;(\langle\ldots h_{\{i,i'\}}\!+\!1\ldots h_{\{i',k\}}\!-\!1\ldots\rangle;\vec{v};\vec{d})) \\
+ & \sum_{i\in K}f(\ell;(\langle\ldots h_{\{i,i\}}\!+\!1\ldots h_{\{i,k\}}\!-\!1\ldots\rangle;\vec{v};\vec{d})) \\
+ & \sum_{(i,i')\in J}f(\ell;(\vec{h};\langle\ldots v_{(i,i')}\!+\!1\ldots v_{(k,i')}\!-\!1\ldots\rangle;\vec{d})) \\
+ & \sum_{\{i,i'\}\in I,\{j,j'\}\in I}
\bigl\{f(\ell;(\langle\ldots h_{\{i,i'\}}\!+\!1\ldots h_{\{j,j'\}}\!+\!1\ldots h_{\{i',j'\}}\!-\!1\ldots\rangle;\vec{v};\vec{d})) \\
 & \ \ \ \ \ \ \ \ \ \ \ \ \ \ \ \ \ \ \ \ +f(\ell;(\langle\ldots h_{\{i,i'\}}\!+\!1\ldots h_{\{j,j'\}}\!+\!1\ldots h_{\{i',j\}}\!-\!1\ldots\rangle;\vec{v};\vec{d})) \\
 & \ \ \ \ \ \ \ \ \ \ \ \ \ \ \ \ \ \ \ \ +f(\ell;(\langle\ldots h_{\{i,i'\}}\!+\!1\ldots h_{\{j,j'\}}\!+\!1\ldots h_{\{i,j'\}}\!-\!1\ldots\rangle;\vec{v};\vec{d})) \\
 & \ \ \ \ \ \ \ \ \ \ \ \ \ \ \ \ \ \ \ \ +f(\ell;(\langle\ldots h_{\{i,i'\}}\!+\!1\ldots h_{\{j,j'\}}\!+\!1\ldots h_{\{i,j\}}\!-\!1\ldots\rangle;\vec{v};\vec{d}))\bigr\}/2 \\
+ & \sum_{i\in K}f(\ell;(\langle\ldots h_{\{i,i\}}\!+\!1\ldots h_{\{j,j'\}}\!+\!1\ldots h_{\{i,j\}}\!-\!1\ldots\rangle;\vec{v};\vec{d})) \\
+ & \sum_{i\in K}f(\ell;(\langle\ldots h_{\{i,i\}}\!+\!1\ldots h_{\{j,j'\}}\!+\!1\ldots h_{\{i,j'\}}\!-\!1\ldots\rangle;\vec{v};\vec{d})) \\
+ & \sum_{i\in K,j\in K}f(\ell;(\langle\ldots h_{\{i,i\}}\!+\!\ldots h_{\{j,j\}}\!+\!1\ldots h_{\{i,j\}}\!-\!1\ldots\rangle;\vec{v};\vec{d})) \\
+ & \sum_{\{i,i'\}\in I,(j,j')\in J}f(\ell;(\langle\ldots h_{\{i,i'\}}\!+\!1\ldots \rangle;\langle\ldots v_{(j,j')}\!+\!1\ldots v_{(i,j')}\!-\!1\ldots \rangle;\vec{d})) \\
+ & \sum_{\{i,i'\}\in I,(j,j')\in J^-}f(\ell;(\langle\ldots h_{\{i,i'\}}\!+\!1\ldots \rangle;\langle\ldots v_{(j,j')}\!+\!1\ldots v_{(i',j')}\!-\!1\ldots \rangle;\vec{d})) \\
+ & \sum_{i\in K,(j,j')\in J}f(\ell;(\langle\ldots h_{\{i,i\}}\!+\!1\ldots \rangle;\langle\ldots v_{(j,j')}\!+\!1\ldots v_{(i,j')}\!-\!1\ldots \rangle;\vec{d})) \\
+ & \sum_{(i,i')\in J,(j,j')\in J}\bigl\{f(\ell;(\vec{h};\langle\ldots v_{(i,i')}\!+\!1\ldots v_{(j,j')}\!+\!1\ldots \rangle;\langle\ldots d_{\{i',j'\}}\!-\!1\ldots \rangle))\bigr\}/2 \\
+ & \sum_{i\in K}f(\ell;(\langle\ldots h_{\{i,k\}}\!-\!1\ldots \rangle;\langle\ldots v_{(i,k)}\!+\!1\ldots \rangle;\vec{d})) \\
+ & \sum_{i\in K}f(\ell;(\vec{h};\langle\ldots v_{(k,i)}\!-\!1\ldots \rangle;\langle\ldots d_{\{i,k\}}\!+\!1\ldots \rangle)) \\
+ & \sum_{i\in K,j\in K}f(\ell;(\langle\ldots h_{\{i,j\}}\!-\!1\ldots \rangle;\langle\ldots v_{(i,k)}\!+\!1\ldots v_{(j,k)}\!+\!1\ldots\rangle;\vec{d}))/2 \\
+ & \sum_{i\in K,j\in K}f(\ell;(\vec{h};\langle\ldots v_{(i,k)}\!+\!1\ldots v_{(i,j)}\!-\!1\ldots\rangle;\langle\ldots d_{\{j,k\}}\!+\!1\ldots\rangle)) \\
+ & \sum_{i\in K,j\in K}f(\ell;(\vec{h};\vec{v};\langle\ldots d_{\{i,k\}}\!+\!1\ldots d_{\{j,k\}}\!+\!1\ldots d_{\{i,j\}}\!-\!1\rangle))/2 \\
+ & \sum_{\{i,i'\}\in I,\{j,k\}\in I}f(\ell;(\langle\ldots h_{\{i,i'\}}\!+\!1\ldots \rangle;\langle\ldots v_{(i',j)}\!-\!1\ldots \rangle;\langle\ldots d_{\{j,k\}}\!+\!1\ldots\rangle)) \\ 
+ & \sum_{\{i,i'\}\in I,\{j,k\}\in I}f(\ell;(\langle\ldots h_{\{i,i'\}}\!+\!1\ldots \rangle;\langle\ldots v_{(i,j)}\!-\!1\ldots \rangle;\langle\ldots d_{\{j,k\}}\!+\!1\ldots\rangle)) \\
+ & \sum_{i\in K,\{j,k\}\in I}f(\ell;(\langle\ldots h_{\{i,i\}}\!+\!1\ldots \rangle;\langle\ldots v_{(i,j)}\!-\!1\ldots \rangle;\langle\ldots d_{\{j,k\}}\!+\!1\ldots\rangle)) \\
+ & \sum_{(i,i')\in J,j\in K}f(\ell;(\vec{h};\langle\ldots v_{(i,i')}\!+\!1\ldots \rangle;\langle\ldots d_{\{j,k\}}\!+\!1\ldots d_{\{i',j\}}\!-\!1\ldots\rangle)). 
\end{array}
\]

\paragraph{Timing analysis} 
First we count the possible combinations of arguments for~$f$. 
Because $\ell$ varies from $0$ to $n$, there are $\Theta(n)$ possible values. 
All of $\vec{h}$, $\vec{v}$, and $\vec{d}$ have $\Theta(c^2)$ elements, 
each of which can have $O(n)$ possible values, 
and therefore $n^{O(c^2)}$ possible values in all. 
Computing a single value of $f$
requires $O(n^4)$ lookups of previously computed values of $f$ 
in case (c), or $O(n^{3c^2})\cdot O(n^2)$ lookups and check-sums 
in cases (b1) and (b2).  The latter bound exceeds the former.
Therefore, the total running time for this DP is 
$\Theta(n)\cdot O(n^{3c^2})\times n^{O(c^2)}
=n^{O(c^2)}$, 
which is polynomial in $n$ when $c$ is a constant. 

\medskip
Because the role of colors and numbers are symmetric in UNO games, 
we have the following result.

\begin{theorem} \label{Uno-1 P}
{\sc Uno-1} is in P if $b$ (the number of numbers) 
or $c$ (the number of colors) is a constant. 
\end{theorem}

\section{Uncooperative UNO}

In this section, 
we study the uncooperative version of UNO. 
\refA{In particular, we show that it is tractable for the two-player case. 
This result is surprising as it contrasts greatly with the cooperative case,
which is intractable for both one and two players.}

\refA{
Let us start our discussions by considering the well-known two-player game {\sc Geography}.} 

\begin{quote}
\refA{{\sc Geography}}
\begin{description}
\item[Instance:] a \refA{directed/undirected} graph, 
and a token placed on an initial vertex.
\item[Question:]
In each turn, the token can be moved to an adjacent vertex, and the \refA{edge/vertex} that
was traversed gets removed from the graph. 
Player 1 and 2 take turns, and the first player unable to move loses. 
Determine the loser.
\end{description}
\end{quote}

\refA{There are in fact four versions of the game,
depending on whether the graph is directed or undirected, 
and whether the traversed vertex or edge gets removed from the graph.
Among those, {\sc Directed Edge Geography} \cite{S78}, 
{\sc Directed Vertex Geography} \cite{LS80} 
and {\sc Undirected Edge Geography} \cite{FSU93} 
are known to be PSPACE-complete. 
Furthermore, {\sc Undirected Edge Geography} remains PSPACE-complete 
even when the graph is planar, and {\sc Directed Edge/Vertex Geography}
remains PSPACE-complete even when the graph is planar or bipartite. 
On the other hand, the following result is crucial in the sense 
that it reveals the difference in complexity 
between {\sc Undirected Vertex Geography} 
and the other three versions.} 

\begin{theorem}{\rm \cite{FSU93}}
\label{uvg_tractable}
\refA{
{\sc Undirected Vertex Geography} is in P. 
}
\end{theorem}

\refA{
This result follows by characterizing solutions to the problem by maximum
matchings.
}

\begin{figure}[hbt]
\centering
\scalebox{0.75}{\input{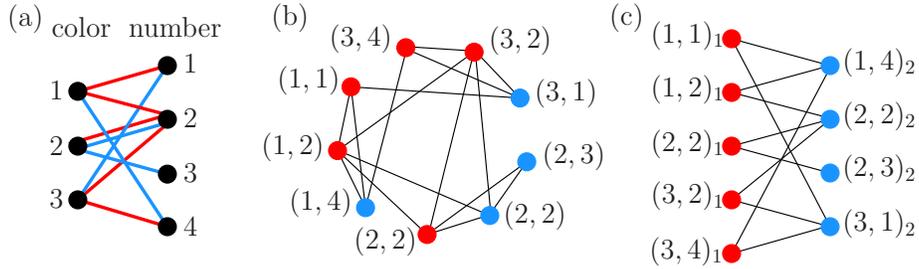}}
\caption{(a) UNO cards are edges of a bipartite graph, 
and they are partitioned into two subsets (red and blue) in UNO-2; 
(b) adjacency of UNO cards becomes a line graph of a bipartite graph; and 
(c) an UNO-2 graph is a bipartite graph by removing edges in each partite set.} 
\label{line_bipartite}
\end{figure}

\refA{
Now we show that {\sc Uncooperative Uno-2} is tractable 
by reducing it to {\sc Undirected Vertex Geography}. 
As we observed in Section \ref{sec:cooperative}, 
because UNO cards can be regarded as edges of a bipartite graph 
(Figure~\ref{line_bipartite}(a)), 
their adjacency (match) relation can be represented by 
a line graph of a bipartite graph (Figure~\ref{line_bipartite}(b)). 
On the other hand, because two players alternate play in UNO-2, 
an UNO-2 graph must be bipartite (and undirected). 
That is, an UNO-2 graph is bipartite and a line graph of a bipartite graph 
(Figure~\ref{line_bipartite}(c)). 
}

\refA{
Then it is not difficult to see that 
{\sc Uncooperative Uno-2} is equivalent 
to {\sc Undirected Vertex Geography} on UNO-2 graphs. 
That is, player 1 wins (is the last player to play cards) 
in {\sc Uncooperative Uno-2}
if and only if player 2 loses (cannot move the token any more) 
in {\sc Undirected Vertex Geography} 
on the corresponding UNO-2 graph. 
Thus, as a corollary of Theorem \ref{uvg_tractable}, 
we have the following result. 
}

\refA{
\begin{theorem}
\label{uncoopUNO2isP}
{\sc Uncooperative Uno-2} is in P. 
\qed
\end{theorem}
}

\medskip
\refA{
We remark that uncooperative UNO is a quite rare example
of a game for which the two-player version is in P
even though the single-player version is NP-complete;
contrast with \cite{Robertson-Munro-1978}.
}

\section{Concluding Remarks}
\label{sec:conclusion}

In this paper, we studied UNO, the well-known card game, 
and gave two mathematical models:
cooperative (to make a specified player win)
and uncooperative (to decide the first player unable to play). 
We analyzed the computational complexity of UNO in several scenarios.
We showed that several variants of these problems are difficult,
but a restricted single-player version are solvable in polynomial time.
On the contrary, we showed that the uncooperative two-player version is in P. 
This is somehow surprising in the sense that multi-players' version 
usually become more intractable than single player's one in many games. 

For future work, it would be interesting to
find a more efficient dynamic program for {\sc Uno-1} 
with a constant number of colors,
by better utilizing its geometric properties. 
In this direction, 
it seems natural to ask whether {\sc Uno-1} 
is fixed-parameter tractable. 
Another direction is to investigate UNO-1 graphs 
from the structural point of view, 
because they form a subclass of claw-free graphs 
and seem to have interesting properties by themselves. 
It also seems promising to make our models more realistic, 
e.g., to take draw pile into account (as an additional player), 
to make all players' cards not open, and so on. 
Finally, the complexity of {\sc Cooperative Uno-2} remains open
when restricted to a constant number of colors.

Based on our mathematical models, 
it is not difficult to invent several variations or generalizations 
of UNO games, even for {\sc Uno-1} (the single-player version). 
Among them, one could generalize an UNO card from a 2-tuple (2-dimensional) 
to $d$-tuple, i.e., {\sc D-dimensional Uno-1} 
with appropriate modifications to the `match' relation of cards. 
Another variation is {\sc Minimum Card Fill-in}, that is, 
given a no instance for {\sc Uno-1}, 
find the minimum number of cards to be added to make it 
to be a yes instance.

\refA{
\section*{Acknowledgments}
We would like to thank the anonymous referees 
for their careful reading and comments 
which greatly helped to correct and improve this paper. 
}

\newpage 
\appendix
\pagenumbering{Roman}
\markboth{APPENDIX}{APPENDIX}

\section[Rules of UNO]{Rules of UNO\protect{\textsuperscript{\textregistered}}}
\label{rule_of_uno}

UNO\textsuperscript{\textregistered} 
is a registered trademark owned by Mattel, Inc. 
We provide here an excerpt of the original rules of UNO. 

\paragraph{Contents}
108 cards as follows: 
\begin{itemize}
\item
19 Blue cards, 0 to 9
\item
19 Green cards, 0 to 9
\item
19 Red cards, 0 to 9
\item
19 Yellow cards, 0 to 9
\item
8 Draw two cards, 2 each in blue, green, red and yellow
\item
8 Reverse cards, 2 each in blue, green, red and yellow
\item
8 Skip cards, 2 each in blue, green, red and yellow
\item
4 Wild cards
\item
4 Wild Draw Four cards
\end{itemize}

\paragraph{Objective of the game}
To be the first player to score 500 points. 
Points are scored by getting rid of all the cards in your hand 
before your opponent(s). 
You score points for cards left in your opponents' hands. 

\paragraph{How to play} 
Every player picks a card. 
The person who picks the highest number deals. 
Action Cards count as zero for this part of the game. 

Once the cards are shuffled each player is dealt 7 cards. 

The remainder of the deck is placed face down to form a DRAW pile. 
The top card of the DRAW pile is turned over to begin a DISCARD pile. 
If an Action Card is the first one turned up from the DRAW pile, 
certain rules apply (see Functions of Action Cards). 

The person to the left of the dealer starts play. 
He/she has to match the card on the DISCARD pile, 
by number, color, or symbol. 
For example, if the card is a \emph{red 7}, 
the player must put down a \emph{red} card or any color \emph{7}. 
Alternatively, 
the player can put down a \emph{Wild} card (see Functions of Action Cards). 

If the player doesn't have a card to match the one on the DISCARD pile, 
he/she \emph{must} take a card from the DRAW pile. 
If the card picked up can be played, the player is free to put it down 
in the same turn. 
Otherwise, play moves on to the next person in turn. 

Players may choose \emph{not} to play a playable card from their hand. 
If so, the player \emph{must} draw a card from the DRAW pile. 
If playable, that card can be put down in the same turn, 
but the player may not use a card from the hand after the draw. 

\paragraph{Functions of the Action Cards} 
The functions of the Action Cards, and when they may be played, 
are set out below. 

\begin{itemize}
\item
{\bf Draw Two Card:}
When this card is played, the next person to play \emph{must draw 2 cards}  
and \emph{miss} his/her turn. 
This card can only be played on matching colors and other Draw Two cards. 
If turned up at the beginning of play, the same rule applies. 
\item
{\bf Reverse Card:}
This simply reverses direction of play. 
Play to changes direction to the right, and vice versa. 
The card may only be played on a matching color or on another reverse card. 
If this card is turned up at the beginning of play, 
the dealer goes first, then play moves to the right instead of the left. 
\item
{\bf Skip Card:}
The next player to play after this card has been laid loses his/her turn 
and is ``\emph{skipped}''. 
The card may only be played on a matching color or on another Skip card. 
If a Skip card is turned up at the beginning of play, 
the player to the left of the dealer is skipped, 
hence the player to the left of that player commences play. 
\item
{\bf Wild Card:}
The person playing this card calls for any color to continue the play, 
including the one currently being played, if so desired. 
A Wild card can be played at any time---even if the player 
has another playable card in the hand. 
If a Wild card is turned up at the beginning of play, 
the person to the left of the dealer determines the color, 
which continues play. 
\item
{\bf Wild Draw Four Card:}
This is the best card to have. 
The person who plays it calls the \emph{color} that continues play. 
Also, the next player has to \emph{pick up 4 cards} from the DRAW pile 
and \emph{miss his/her turn}. 
Unfortunately, the card can only be played when the player holding it 
does not have a card in his/her hand to match the \emph{color} 
on the DISCARD pile. 
If the player holds matching numbers or Action Cards, however, 
the Wild Draw Four card may be played. 
A player holding a Wild Draw Four may choose to bluff and play the card 
illegally, but if he/she is \emph{caught} certain rules apply. 
If this card is turned up at the beginning of play, 
it is returned to the deck and another card is picked. 
\end{itemize}

\paragraph{Going out} 
When a player has only one card left, he/she must yell ``\emph{UNO}''. 
Failure to do this results in having to pick up 2 cards from the DRAW pile. 
This is only necessary, however, if he/she is caught by one of 
the other players. 

Once a player has no cards left, the hand is over. 
Points are scored and play starts over again. 

If the last card played in a hand is a Draw Two of Wild Draw Four card, 
the next player \emph{must} draw the 2 or 4 cards respectively. 
These cards are counted when the points are totalled. 

If no player is out of cards by the time the DRAW pile is depleted, 
the deck is reshuffled and play continues.

\end{document}